\documentclass[aps,prl,twocolumn,superscriptaddress]{revtex4}

\usepackage{amsmath}
\usepackage{amsfonts}
\usepackage{graphicx}

\begin{document}


\title{Nonlocal Optics of Plasmonic Nanowire Metamaterials}


\author{Brian M. Wells}
\affiliation{Department of Physics and Applied Physics, University of Massachusetts Lowell, Lowell, MA, 01854}

\author{Anatoly V. Zayats}
\affiliation{Department of Physics, King's College London, Strand, London, WC2R 2LS, UK}

\author{Viktor A. Podolskiy}
\affiliation{Department of Physics and Applied Physics, University of Massachusetts Lowell, Lowell, MA, 01854}


\date{\today}

\begin{abstract}
We present an analytical description of the nonlocal optical response of plasmonic nanowire metamaterials that enable negative refraction, subwavelength light manipulation, and emission lifetime engineering. We show that dispersion of optical waves propagating in nanowire media results from coupling of transverse and longitudinal electromagnetic modes supported by the composite and derive the nonlocal effective medium approximation for this dispersion. We derive the profiles of electric field across the unit cell, and use these expressions to solve the long-standing problem of additional boundary conditions in calculations of transmission and reflection of waves by nonlocal nanowire media. We verify our analytical results with numerical solutions of Maxwell's equations and discuss generalization of the developed formalism to other uniaxial metamaterials.
\end{abstract}

\pacs{78.20.Ci, 42.70.-a, 78.20.Bh}

\maketitle

Nanowire-based composites have recently attracted significant attention due to their unusual and counterintuitive optical properties that include negative refraction, subwavelength confinement of optical radiation, and modulation of photonic density of states\cite{kivsharReview,nrefractive,dstates}. Due to relatively low loss and ease of fabrication, nanowire composites found numerous applications in subwavelength imaging, biosensing, acousto-optics, and ultrafast all-optical processing, spanning visible to THz frequencies\cite{subimag,hypergrat,hyperlens,biosensor,cloak}. Wire materials are a sub-class of uniaxial metamaterials that have homogeneous internal structure along one pre-selected direction. In all, this class of composites represents a flexible platform for engineering of optical landscape from all-dielectric birefringent media, to epsilon-near-zero, to hyperbolic, and epsilon-near-infinity regimes, each of which has its own class of benefits and applications\cite{ENZ,metaBook}.

In this work, we present an analytical technique that provides adequate description of electromagnetism in wire-based metamaterials taking into account nonlocal optical response originating from the homogenization procedure. The approach can be straightforwardly extended to describe optics of coaxial-cable-like media\cite{coax} and numerous other uniaxial composites. The developed formalism reconciles the local and nonlocal effective-medium theories often used to describe the optics of nanowire composites in different limits\cite{nwEMT,PECwires,efros,podolskiy}. More importantly, the formalism relates the origin of optical nonlocality to collective (averaged over many nanowires) plasmonic excitation of wire composite, and provides the recipe to implement additional boundary conditions in composite structures. 

We illustrate the developed technique on the example of plasmonic nanowire metamaterials, formed by an array of aligned plasmonic nanowires embedded in a dielectric host. For simplicity, we fix the frequency of electromagnetic excitations and the unit cell parameters of the system, and vary only the permittivity of the wire inclusions. (The developed formalism can be readily applied for systems where both permittivity and frequency are changed at the same time.) We assume that the system operates in the effective-medium regime (its unit cell $a\ll\lambda_0$ with $\lambda_0$ being the free-space wavelength) and that the surface concentration of plasmonic wires is small $p=\pi R^2/a^2\ll1$. The parameters used in this work are $R=20 $nm, $a=100 $nm, $\epsilon_h$=1, L =1 $\mu$m, $\lambda_0$=1.5 $\mu$m (see Fig.1), which are typical for composites fabricated with anodized alumina templates\cite{nrefractive,biosensor}. 

As previously mentioned, the optical response of nanowire materials resembles that of uniaxial media with optical axis parallel to the direction of the nanowires ($z$). Therefore, dielectric permittivity tensor describing properties of waves propagating in the wire media is diagonal with components $\epsilon_x=\epsilon_y=\epsilon_{x,y}$ and $\epsilon_z$.

\begin{figure}[b]
\includegraphics[width=9cm]{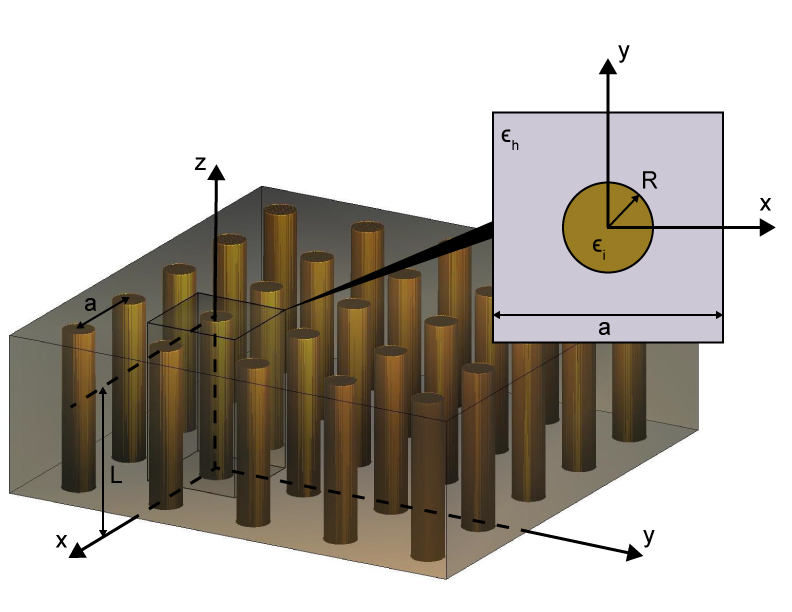}%
\caption{\label{fig:Figure1} Schematic geometry and a unit cell of a nanowire composite.
}
\end{figure}

It has been shown that at optical and near-IR frequencies, the behavior of these components is largely described by Maxwell-Garnett type effective medium theory (EMT)\cite{MG,nwEMT}. In this approach, the microscopic distribution of the field is given by  solutions of the Maxwell equations in quasistatic limit
\begin{eqnarray}
\label{eqMGfld}
E_z^{mg}&=&\mathfrak{e}_{z}^{mg}
\\
\nonumber
E_{x}^{mg}&=&\mathfrak{e}_{x}^{mg}  \times\left\{
\begin{array}{l}
2\frac{\epsilon_h}{\epsilon_h+\epsilon_i},r\le R
\\
1+R^2\frac{\epsilon_h-\epsilon_i}{\epsilon_h+\epsilon_i}\frac{y^2-x^2}{(x^2+y^2)^2},r>R
\end{array}
\right.
\end{eqnarray}
with $\epsilon_i$ and $\epsilon_h$ being the permittivities of wire inclusions and of host material respectively, and parameters $\mathfrak{e}_{x}^{mg}$ and $\mathfrak{e}_{x}^{mg}$ are the field amplitudes. Straightforward averaging of the $j$-th component of the fields over the unit cell yields the effective permittivity $\epsilon_j^{mg}=\langle \epsilon(x,y) E_j^{mg}(x,y)\rangle/\langle E_j^{mg}(x,y)\rangle$: 
\begin{equation}
\label{eqLocalEMT}
\begin{aligned}
\epsilon_{x,y}^{mg}&=\epsilon_h\frac{(1+p)\epsilon_i+(1-p)\epsilon_h}{(1+p)\epsilon_h+(1-p)\epsilon_i}\\
\epsilon_z^{mg}&=p\epsilon_i+(1-p)\epsilon_h
\end{aligned}
\end{equation}

By adjusting composition of the metamaterial and operating wavelength, the optical response of the composite can be controlled between all-dielectric elliptic ($\epsilon^{mg}_{x,y}>0,\epsilon^{mg}_z>0$), epsilon-near-zero (ENZ, $\epsilon^{mg}_z\simeq 0$) and hyperbolic ($\epsilon^{mg}_{x,y}>0,\epsilon^{mg}_z<0$) regimes. In the two latter regimes, metamaterial supports optical waves with either small or large effective modal index and motivate a number of potential applications in molding of light\cite{ENZ}, cloaking\cite{cloak}, and subwavelength light manipulation\cite{hypergrat,hyperlens,kivsharReview}.

At the same time, it has been shown that at lower frequencies where $-\epsilon_i\gg 1$,  $\epsilon_z$ of wire composites becomes strongly nonlocal (exhibits strong dependence on $k_z$)\cite{PECwires,efros}. Similar dependence has been recently shown to take place at visible frequencies in the ENZ regime\cite{podolskiy}. Nonlocality, especially, in the ENZ regime, has been shown to fundamentally alter the optical response of wire composite, leading to excitation of new types of optical waves, and requiring additional boundary conditions for their analytical description\cite{pekar, podolskiy,PECwires}. Despite extensive previous research, existing first-principal theoretical models describing optics of wire composites\cite{PECwires} cannot be used at visible and [near]-IR frequencies, with remaining models requiring fitting\cite{podolskiy} or numerical solutions of Maxwell equations. Here we present a formalism free of the above shortcomings.

The dispersion of plane waves propagating inside homogeneous (nonlocal) uniaxial materials can be derived from the well-known relation
\begin{equation}
\label{eqDispMatrix}
det\left| \vec{k}\cdot\vec{k}\;\delta_{ij}-k_i k_j-\epsilon_{ij}\frac{\omega^2}{c^2}\right|=0
\end{equation}
with $\vec{k}$ and $\omega$ being the wavevector and the angular frequency of the plane wave, respectively, $c$ being the speed of light in vacuum, $\epsilon$ being the generally nonlocal dielectric permittivity tensor of the metamaterial,  and subscripts $ij$ correspond to Cartesian components\cite{jackson}. In the local regime, $\epsilon_{ij}=\epsilon_{ij}^{mg}$.

We now focus on the development of the model for the nonlocal effective permittivity of a nanowire composite.
For propagation along the optical axis ($k_x=k_y=0$), Eq.(\ref{eqDispMatrix}) allows two types of solutions. One is the well-known solution $k_z^{mg}=\sqrt{\epsilon_{x,y}}\omega/c$. In nanowire composites, this solution corresponds to the local permittivity $\epsilon_{x,y}=\epsilon_{x,y}^{mg}$ which is related to plasmonic oscillations perpendicular to the wire axes\cite{nwEMT} and is thus influenced by the plasmonic resonance of the composite, slightly shifted from the position of the localised surface plasmon resonances in isolated wires due to inter-wire interaction.

The second solution of Eq.(\ref{eqDispMatrix}) corresponds to $\epsilon_z(k_z)=0$. This describes a longitudinal wave propagating in z-direction, with $\vec{E}\|\vec{k}\|\hat{z}$. This solution is also known as the additional (TM-polarized) wave. 

As we show below, in nanowire systems this mode results from the interaction between cylindrical surface plasmon (CSP) modes\cite{boardmanBook} of the individual wires that comprise the collective (related to many wires) longitudinal electromagnetic mode. The components of the fields of the this mode can be related to its $z$-components, that in turn can be written as a linear combination of cylindrical waves. For the square unit cell geometry, considered in this work, the latter combination will only contain cylindrical modes with $m=0,4,8,\ldots$. Explicitly,
\begin{widetext}
\begin{eqnarray}
\label{eqDispSeries}
E_z^l&=&e^{-ik_z^lz}\sum_m\cos(m \phi)\times
\left\{
\begin{array}{l}
a_m J_m(\kappa_i r), r\le R
\\
\lbrack \alpha^+_m H^+_m(\kappa_h r)+ \alpha^-_m H^-_m(\kappa_h r)\rbrack,r>R
\end{array}
\right.
\\
\nonumber
H_z^l&=&e^{-ik_z^lz}\sum_m
\left\{
\begin{array}{l}
1,m=0
\\
\sin(m \phi),m\ge 1
\end{array}
\right\}
\times
\left\{
\begin{array}{l}
b_m J_m(\kappa_i r), r\le R
\\
\lbrack \beta^+_m H^+_m(\kappa_h r)+ \beta^-_m H^-_m(\kappa_h r)\rbrack,r>R
\end{array}
\right.
\end{eqnarray}
\end{widetext}
with $\kappa_{\{i,h\}}^2+{k_z^l}^2=\epsilon_{\{i,h\}}\omega^2/c^2$. Note that continuity of $E_z, H_z, E_\phi$, and $H_\phi$ at $r=R$ uniquely define the parameters $\alpha_m^+,\beta_m^+,a_m$, and $b_m$ as a function of $\alpha_m^-$, and $\beta_m^-$. 

\begin{figure}[b]
\includegraphics[width=8cm]{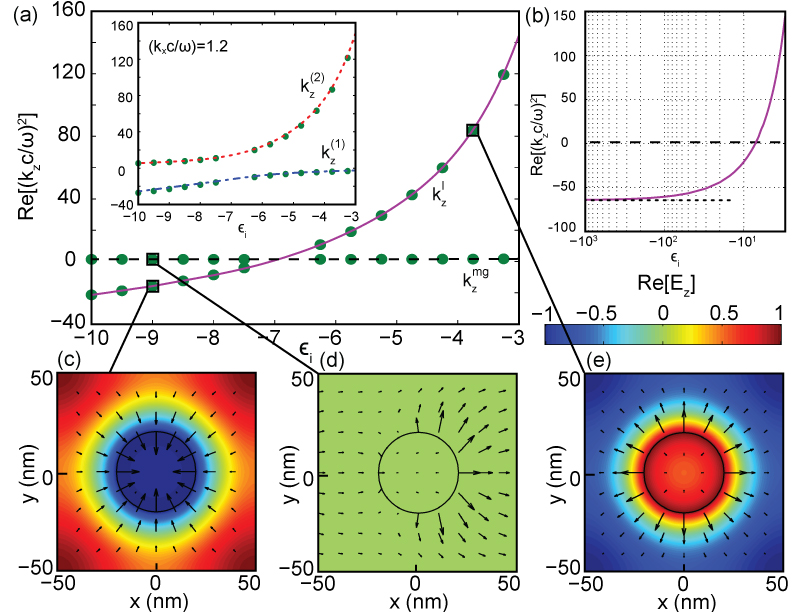}
\caption{\label{Figure2} (a,b) Dispersion in nanowire composite as a function of wire permittivity. Dashed and solid lines represent transverse and longitudinal waves [$k_z^{mg}$ and $k_z^l$,Eq.(\ref{eqDispEqL})] respectively; inset in (a) represents off-axis propagation [Eq.(\ref{eqDispOscillator})]; symbols represent numerical solutions of the Maxwell's equations; for  $-\epsilon_i\gg \epsilon_i$, $k_z^l \rightarrow n^l_{\infty}\omega/c$ [dotted line in (b)]. (c,d,e) electric field in the unit cell; surface plots and arrows represent $E_z$ and $\vec{E}_{x,y}$ components, respectively.
}
\end{figure}

The field of the mode has to satisfy the periodicity condition $E,H|_{x=-a/2}=E,H|_{x=+a/2}$. An analysis of the fields produced by a series in Eq.(\ref{eqDispSeries}) suggests that such a solution can be realized when $\beta_m^-=0$, and $k_z$ and $\alpha_m^-$ are obtained from the eigenvalue-type problem
\begin{eqnarray}
\label{eqDispEqL}
\sum_m\widehat{\mathcal{H}_y}_{jm}(k_z^l)\alpha_m^-=0
\end{eqnarray}
with $jm$ elements of matrix $\widehat{\mathcal{H}_y}$ corresponding to the $y$-component of the magnetic field produced by the $m$-th Hankel function at the location $\{x,y\}=\left\{\frac{a}{2},j\frac{a}{2N_{m}}\right\}$ with $N_m$ being the number of $m$ terms in Eq.(\ref{eqDispSeries}). Our analysis suggests that as number of terms in Eq.(\ref{eqDispSeries}) increases, the dispersion produced by the analytical technique quickly converges to the dispersion obtained by direct numerical solution of Maxwell's equations (here we use finite element method and rigorous-coupled-wave analysis). Fig.2 demonstrates the excellent agreement between the numerical and analytical solutions corresponding to a three-term series $m=[0,4,8]$, and clearly demonstrates the longitudinal character of this mode.
This wave is strongly dispersive in the regime $\epsilon_i \rightarrow - \epsilon_h$, corresponding to the surface plasmon oscillations on the metal-dielectric interface. On the other hand, when $-\epsilon_i\gg \epsilon_h$ (realized at mid-IR and lower frequencies for noble metals), the wavevector of the longitudinal mode approaches $n^l_\infty\omega/c$, and its transverse counterpart approaches light line(see Ref.\cite{PECwires} and Fig.2).

Comparing the dispersion relation corresponding to microscopic [Eq.(\ref{eqDispEqL})] and effective medium approximation $\epsilon_z(k_z)=0$, a complete description of the nonlocal effective permittivity can be obtained
\begin{equation}
\label{eqEz}
\epsilon_z(k_{z})=\xi \left(k_z^2-{k_z^l}^2\right)\frac{c^2}{\omega^2}
\end{equation}
where $k_z$ is the wavevector of the mode in the nonlocal effective medium approximation, $k_z^l$ is the wavevector of the mode of the composite in the microscopic theory, and $\xi$ is the factor which will be determined below.

The above considerations can be extended to the case of propagation of waves at an angle to the optical axis. For simplicity, we consider the case $k_y=0,k_x\neq 0$. It is relatively straightforward to transform Eq.(\ref{eqDispMatrix}) into a set of two uncoupled dispersion relations. For nanowire composites, the first of these, $k_x^2+k_z^2=\epsilon_{x,y}^{mg}\omega^2/c^2$ describes the propagation of transverse-electric (TE)-polarized waves. The second,
$\epsilon_z(k_z)\left(k_z^2-\epsilon_{x,y}^{mg}\frac{\omega^2}{c^2}\right)=-\epsilon_{x,y}^{mg} k_x^2$, describes the propagation of the transverse-magnetic (TM)-polarized waves. Taking into account Eq.(\ref{eqEz}), the latter relation can be re-written as
\begin{equation}
\label{eqDispOscillator}
\left(k_z^2-{k_z^l}^2\right)
\left(k_z^2-\epsilon_{x,y}^{mg}\frac{\omega^2}{c^2}\right)=-\frac{\epsilon_{x,y}^{mg}}{\xi} \frac{\omega^2}{c^2}k_x^2
\end{equation}
that reflects the fact that similar to other nonlocal materials\cite{pekar}, nanowire composites support two TM-polarized waves propagating with different indices.

Eq.(\ref{eqDispOscillator}) clearly shows that off-angle ($k_x\neq 0$) propagation of the two TM-polarized waves in anisotropic wire media can be mapped to a microscopic model of optical properties of a nanowire array. In this description, the two TM modes are determined by the components of the effective permittivities arising from (i) transverse (electron oscillations perpendicular to the nanowire axes) and (ii) longitudinal (electron oscillations and the wavevector parallel to the nanowire axes) parts of the cylindrical plasmons of the wires. The off-angle wavevector plays the role of the coupling constant. This nonlocality is present only in the effective medium model due to the homogenization procedure; in the microscopic model of the nanowire array, all the quantities are local.

The remaining free parameter of the model, multiplicative factor $\xi$, can be determined by requiring that in the limit of small $k_x$ the properties of one of the two TM-polarized waves follow elliptic or hyperbolic dispersion and has $k_z(k_x)=\rm const$ dependence, observed in the wire media when $\epsilon_z^{mg}>0, \epsilon_z^{mg}\lesssim 0$, and $\epsilon_z^{mg}\ll -1$ respectively\cite{nwEMT,PECwires}. The relationship
\begin{equation}
\xi=p\frac{\epsilon_i+\epsilon_h}{\epsilon_h-{n_l^\infty}^2 }
\end{equation}
adequately describes optics of wire media in these three limits. The excellent agreement between the predictions of Eq.(\ref{eqDispOscillator}) and the full-wave numerical solutions of Maxwell equations is shown in Figs.(2,3). As expected, the isofrequency of the ``main'' TM-polarized wave resembles ellipse or hyperbola that for small values of $k_x$ is well-described by $\hat{\epsilon}^{mg}$. At the same time, the dependence $k_z(k_x)$ in ``additional'' wave is opposite to that of its ``main'' counterpart. The $z$-component of permittivity is described by Eq.(\ref{eqLocalEMT}) only for on-axis ($k_x=0$) propagation, and exhibits strong spatial dispersion for oblique propagation of light.  

The presented formalism provides a mechanism to calculate the deviation of the dispersion of the waves in nanowire materials from the prediction of local effective medium technique. In particular, this deviation places fundamental limits on subwavelengh light manipulation and on increase of local density of photonics states\cite{dstates}. It is also likely to affect the cloaking performance of nanowire-based structures\cite{cloak}.

\begin{figure}[tb]
\includegraphics[width=9cm]{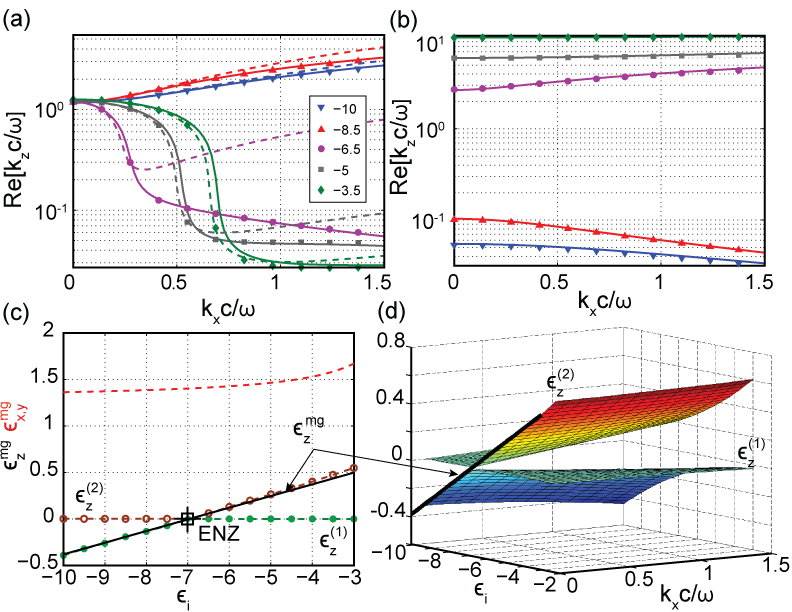}
\caption{\label{Figure3} (a,b) Isofrequency contours of TM-polarized modes in nanowire composites. Solid lines and symbols correspond to Eq.(\ref{eqDispOscillator}) and numerical solutions of Maxwell equations respectively; dashed lines represent local EMT. (c,d) The effective medium permittivity of nonlocal nanowire composite for $k_x=0$ (c) and for $k_x\neq 0$(d); $\epsilon_z^{(1)},\epsilon_z^{(2)}$ represent two solutions of Eq.(\ref{eqEz}); cross marks ENZ condition ($\epsilon_z^{mg}\simeq 0$)
}
\end{figure}

Now that the origin and dispersion of the modes propagating in nanowire systems is understood, we focus on the analysis of the optical properties of finite-size wire arrays. Since in the EMT approximation the fields of TE and TM-polarized modes are orthogonal to each other, and since propagation of TE-polarized light through the wire-based system is only affected by $x,y$ components of permittivity, this propagation can be successfully described by Eq.(\ref{eqLocalEMT})\cite{nwEMT}. Here, we focus on the analysis of propagation of TM-polarized light. This analysis must answer two important questions: (i) what is the structure of electromagnetic waves propagating in the system, and (ii) what are the additional boundary conditions needed to determine the amplitudes of the two TM-polarized modes inside the wire system.

Consistent effective medium description requires that the unit-cell averaged fields satisfy both constituent relations $\epsilon_j=\langle \epsilon E_j\rangle/\langle E_j\rangle$ and relations between the field components of the plane wave $\langle \epsilon E_z\rangle=-k_x/k_z\langle \epsilon E_x\rangle$. With these constraints, we start from Eqs.(\ref{eqMGfld},\ref{eqDispSeries}), determine the parameters $\alpha_0^-,\mathfrak{e}^{mg}_{z}$, and $\mathfrak{e}^{mg}_{x}$ by normalizing $\langle E_z^l \rangle=\langle E_x^{mg}\rangle=\langle E_z^{mg}\rangle=1$, and construct the fields of the two waves propagating in the wire media as $\vec{E}(x,y)e^{i\omega t-i k_x x-ik_zz}$ with
\begin{eqnarray}
\label{eqEmicro}
E_x(x,y)&=&E_x^{mg}
\\
\nonumber
E_z(x,y)&=&\gamma^{mg} E_z^{mg}+\gamma^l \left.E_z^l\right\vert_{z=0}
\end{eqnarray}
and
\begin{eqnarray}
\gamma^{mg}=-\frac{\epsilon_{x,y}^{mg} k_x}{\epsilon_z k_z}\frac{\epsilon_z-\epsilon^l}{\epsilon_{z}^{mg}-\epsilon^l}
\\ \nonumber
\gamma^l=-\frac{\epsilon_{x,y}^{mg} k_x}{\epsilon_z k_z}\frac{\epsilon_z-\epsilon_{z}^{e}}{\epsilon^l-\epsilon_{z}^{e}}
\end{eqnarray}
In the expressions above $\epsilon_z\equiv \epsilon_z(k_z)$ is given by Eq.(\ref{eqEz}), $k_z(k_x)$ -- by Eq.(\ref{eqDispOscillator}), and $\epsilon^l=\langle \epsilon(x,y) E_z^l(x,y)\rangle/\langle E_z^l(x,y)\rangle$.

Eq.(\ref{eqEmicro}) represents a transition between full-wave solutions of Maxwell equations in the nanowire array where the fields oscillate on the scale of the individual wires, and effective-medium solutions where plane waves propagate in the homogenized material. Since our model for $\vec{E}^{mg}$ assumes quasistatic limit [Eq.(\ref{eqMGfld})], Eq.(\ref{eqEmicro}) is technically valid in the limit $a\ll 2\pi/k_x,\lambda_0$. However, Fig. 3 indicates that the developed formalism also provides adequate approximation for optics of wire systems for higher values of $k_x$.

Finally, we consider the problem of reflection/refraction of light at the interface of two (nonlocal) (meta-) materials, extending the well-developed transfer-matrix formalism\cite{TMM} to nonlocal composites. Maxwell equations require continuity of (microscopic) $E_x$ and $D_z$, and the effective-medium boundary conditions can be obtained by averaging these relationships across the unit cell. Multiple, linearly independent boundary conditions can be obtained by requiring the continuity of $\mathbb E^n =\langle e^{2\pi i n \frac{x}{a}} E_x\rangle$ and $\mathbb D^n=\langle e^{2\pi i n\frac{ x}{a}} D_z\rangle$ with different integer $n$.

In particular, for the interface between two conventional materials, continuity of $\mathbb{E}^0,\mathbb{D}^0$ yield conventional TMM results. If one of the media is nonlocal metamaterial, we suggest to complement the above boundary conditions by the additional boundary condition (ABC) of continuity of $\mathbb{D}^l$. Since polarization along the optical axis $P_z$ is dominated by the field of longtitudinal wave, this ABC is close to the heuristic condition $P_\perp^{\rm nonlocal}=0$, proposed for homogeneous\cite{pekar} and nonlocal\cite{podolskiy} materials, and to condition of continuity of $\epsilon_h\mathbb{E}^0$, suggested for ultra-thin high-conductivity wires\cite{PECwires} [note that in general the above ABCs are not identical to each other].
Finally, when both media are nonlocal, continuity of $\mathbb{E}^1$ plays the role of the second additional boundary condition. Note that due to presence of longitudinal modes at both sides of the interface, $P_\perp^{\rm nonlocal}$ may not vanish in this configuration. As an additional cross-check, the proposed combination of ABCs ensures full transmission of light through ``virtual" interface between two identical nanowire metamaterials.

Transmission and reflection of metamaterial are compared for full-vectorial numerical solutions of Maxwell's equations, predictions of conventional (local) effective medium theory, and predictions of the nonlocal EMT developed in this work in Fig.4. It is seen the smaller the loss and the larger the angle of incidence the more important it becomes to take into account the nonlocal optics of nanowire composites. Interestingly, nonlocal response strongly affects optical response of the wire metamaterials across the broad range of the effective permittivities. This effect most clearly seen in reflection, but is also visible in transmission, especially in ENZ regime. 

\begin{figure}[tb]
\includegraphics[width=9cm]{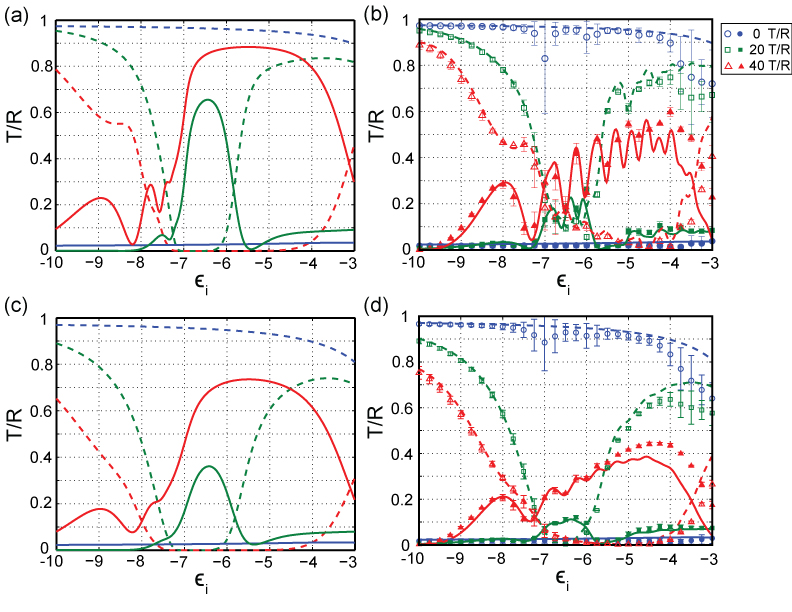}%
\caption{\label{Figure4} Transmission and reflection of light through a parallel slab of nanowire media, suspended in air with $Im(\epsilon_i)=-0.1$ (a,b) and $Im(\epsilon_i)=-0.25$ (c,d). (a,c): local TMM calculations, (b,d): nonlocal EMT developed here (lines) and numerical solutions of Maxwell equations (symbols); Solid lines and filled symbols represent reflection, dashed lines and empty symbols - transmission.
}
\end{figure}

To conclude, we presented an approach to describe optics of nonlocal wire metamaterials across the whole optical spectrum. The formalism demonstrates excellent agreement with the results of numerical solutions of Maxwell's equations and is essential in development of an adequate description of optics in wire arrays, from understanding the true density of photonic states to the limits of resolution in these systems. The developed formalism can be straightforwardly extended to describe the optics of other wire-like composites including coated wire-structures, and coax-cable-based systems\cite{coax}.

This research is supported by US Army Research Office (grant \# W911NF-12-1-0533), EPSRC (UK), and the ERC iPLASMM project (321268).

\end{document}